\documentclass[prl,aps,twocolumn,pdffig]{revtex4}
\usepackage{graphicx}
\usepackage{dcolumn}
\usepackage{bm}
\makeatletter

\newcommand{\Rmnum}[1]{\expandafter\@slowromancap\romannumeral #1@}
\makeatother
\begin{document}
\title {Large low-frequency resistance noise in chemical vapor deposited
graphene}
\author{Atindra Nath Pal,$^1$ Ageeth A. Bol,$^2$ and Arindam Ghosh$^1$ }
\vspace{1.5cm}
\address{$^1$ Department of Physics, Indian Institute of Science, Bangalore 560 012, India}
\address{$^2$ IBM Thomas J. Watson Research Center, Yorktown Heights, NY 10598, USA}
\begin{abstract}
We report a detailed investigation of resistance noise in single layer graphene films on  Si/SiO$_2$ substrates obtained by chemical vapor
deposition (CVD) on copper foils. We find that noise in these systems to be rather large, and when expressed in the form of phenomenological
Hooge equation, it corresponds to Hooge parameter as large as $0.1 - 0.5$. We also find the variation in the noise magnitude with the gate
voltage (or carrier density) and temperature to be surprisingly weak,  which is also unlike the behavior of noise in other forms of graphene, in
particular those from exfoliation.
\end{abstract}


\maketitle

The electronic properties of graphene has recently been the subject of intense research for both fundamental science and technological
applications. Mechanically exfoliated graphene offers the cleanest devices with mobility in the range of $\sim 200,000$
cm$^2$/Vs~\cite{suspended_Kim,{suspended_andrei}}, forming the backbone of fundamental phenomena such as the fractional quantum Hall
effect~\cite{FQHE_andrei,{FQHE_kim}}, or ultra-high frequency transistors~\cite{50GhZ_graphene}. The exfoliation process is however statistical,
and for regular large scale production, several new methods have been suggested including epitaxial growth of graphene on SiC
wafers~\cite{SiC_1,{SiC_2}}, reduction of graphene oxide~\cite{graphite_oxide}, and thermally grown graphene from decomposition of hydrocarbon
(methane) on transition metal (copper, nickel, iridium etc.)
surfaces~\cite{CVD_Nature,{CVD_science},{CVD_1},{CVD_2},{CVD_4},{CVD_5},{CVD_6},{CVD_Park},{APL_QHE}}. The latter metal-based chemical vapor
deposition (CVD) technique of realizing large area graphene is of particular interest as it displays excellent electrical (high
mobility~\cite{CVD_science}, low resistance/square, half-integer quantum Hall effect~\cite{APL_QHE}), mechanical (large gauge factor and
electromechanical stability~\cite{CVD_Nature}) and optical (high transmittance~\cite{CVD_5}) properties. Moreover, recent developments in
transferring large films of single-layer CVD-graphene onto insulating substrates offer great promise in nanoelectronics, transparent electrodes
in solar photovoltaics~\cite{CVD_5}, or flexible/stretchable electronic applications~\cite{CVD_6}. An important aspect of such applications is
the intrinsic electrical noise in CVD-graphene films, which has not been explored so far. A study of electrical noise may also be crucial in
understanding the nature of disorder in these materials which can be significantly different from the other forms of
graphene~\cite{atin_prl,{atin_apl}}. In this letter we report the first experimental investigation of low-frequency fluctuations of electrical
resistance, often known as the $1/f$-noise or flicker noise, in large-area films of single layer graphene (SLG) grown on Cu-foil and
subsequently transferred onto Si/SiO$_2$ substrate. We find the noise in CVD-graphene to be significantly larger than typical exfoliated
graphene devices, along with several surprising features that separates the kinetics of disorder in CVD graphene from other graphene systems.

\begin{figure}[!t]
\begin{center}
\includegraphics [width=8cm,height=6.5cm]{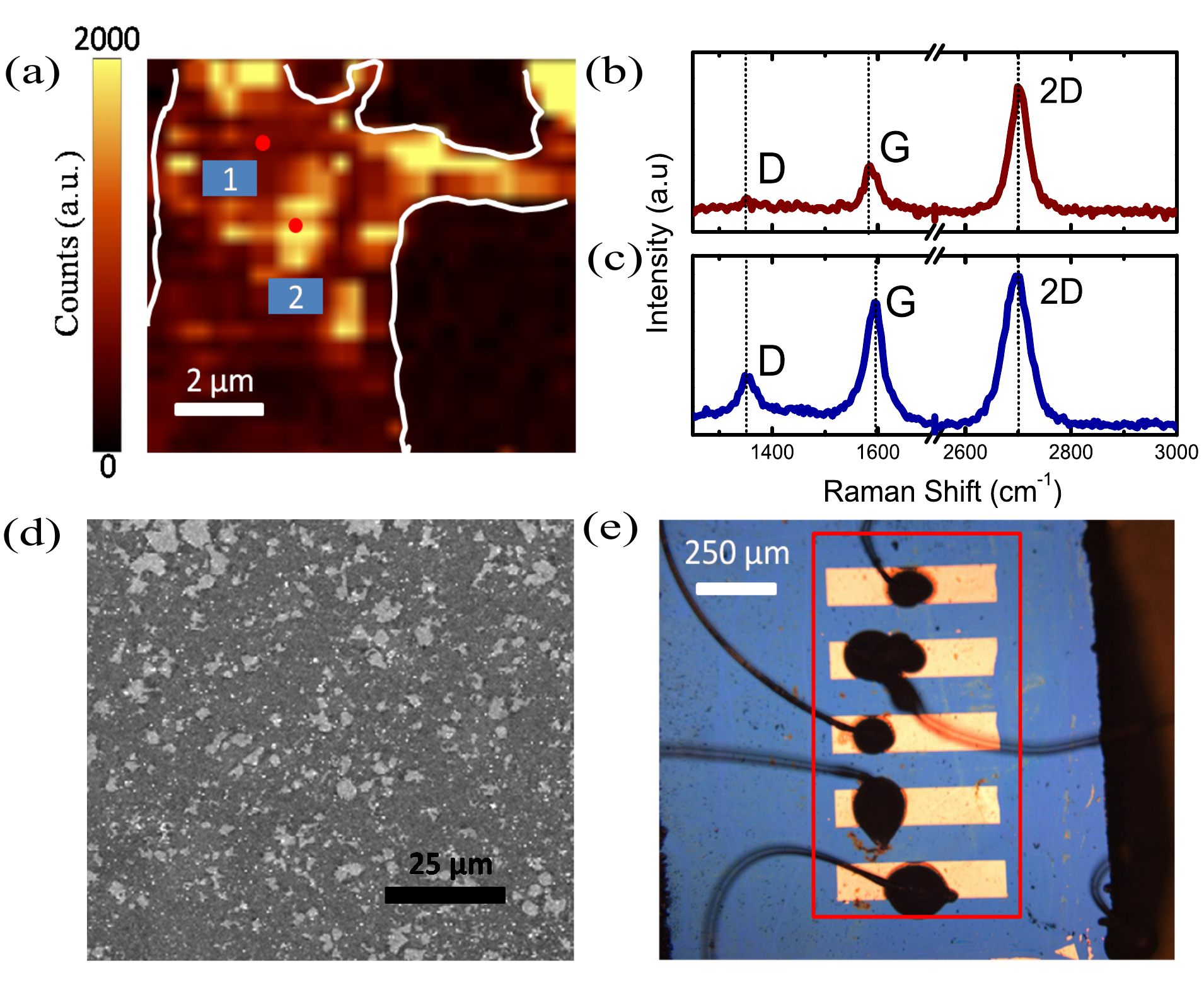}
\end{center}
\caption{Color Online. (a) D peak intensity map of a portion of graphene on SiO$_2$, where the graphene is outlined by thick white line. Raman
spectra of graphene corresponding to the positions 1 and 2, in Fig. 1a, are shown in (b) and (c) respectively. These indicate the spatial
variation of D peak intensity, with varying $I_{2D}/I_G$ ratio. (d) SEM image of CVD graphene indicating the ruptures. (e) Optical micrograph of
a typical device outlined by the rectangle. } \label{figure1}
\end{figure}

Recent studies of carrier mobility ($\mu$)~\cite{Avouris_PRB} and $1/f$
noise~\cite{Avouris,{atin_prl},{atin_apl}} in exfoliated graphene on insulating
substrates indicates that both static (that gives rise to average resistivity)
and time varying (resulting in noise) components of disorder are dominated by
the trapped charges at the graphene-substrate interface. This is particularly
true at low carrier density ($n$) where scattering off the Coulomb potential
from the trapped charges leads to a linear dependence of graphene conductivity
($\sigma \propto n$)~\cite{sdsarma_BLG}. Short range scattering, involving for
examples lattice defects or neutral impurities etc, become important only at
large $n$ where the Coulomb potentials are largely screened. In CVD-graphene
the situation can be very different. The process of etching of the host metal,
mechanical stressing during the transfer process etc., have been shown to lead
to considerable additional disorder, which manifests in lower $\mu$, and often
a clearly visible D-peak in Raman spectroscopy~\cite{CVD_Park}. Indeed, low
temperature magnetoresistance measurements in CVD-graphene reveal a short
elastic intra-valley mean free path, indicating presence of spatially extended
defects, such as line defects, dislocations and ripples~\cite{APL_QHE}. Whether
this additional disorder can also cause higher noise in CVD-graphene is not
known.

\begin{figure}[!tbp]
\begin{center}
\includegraphics [width=7.5cm,height=8cm]{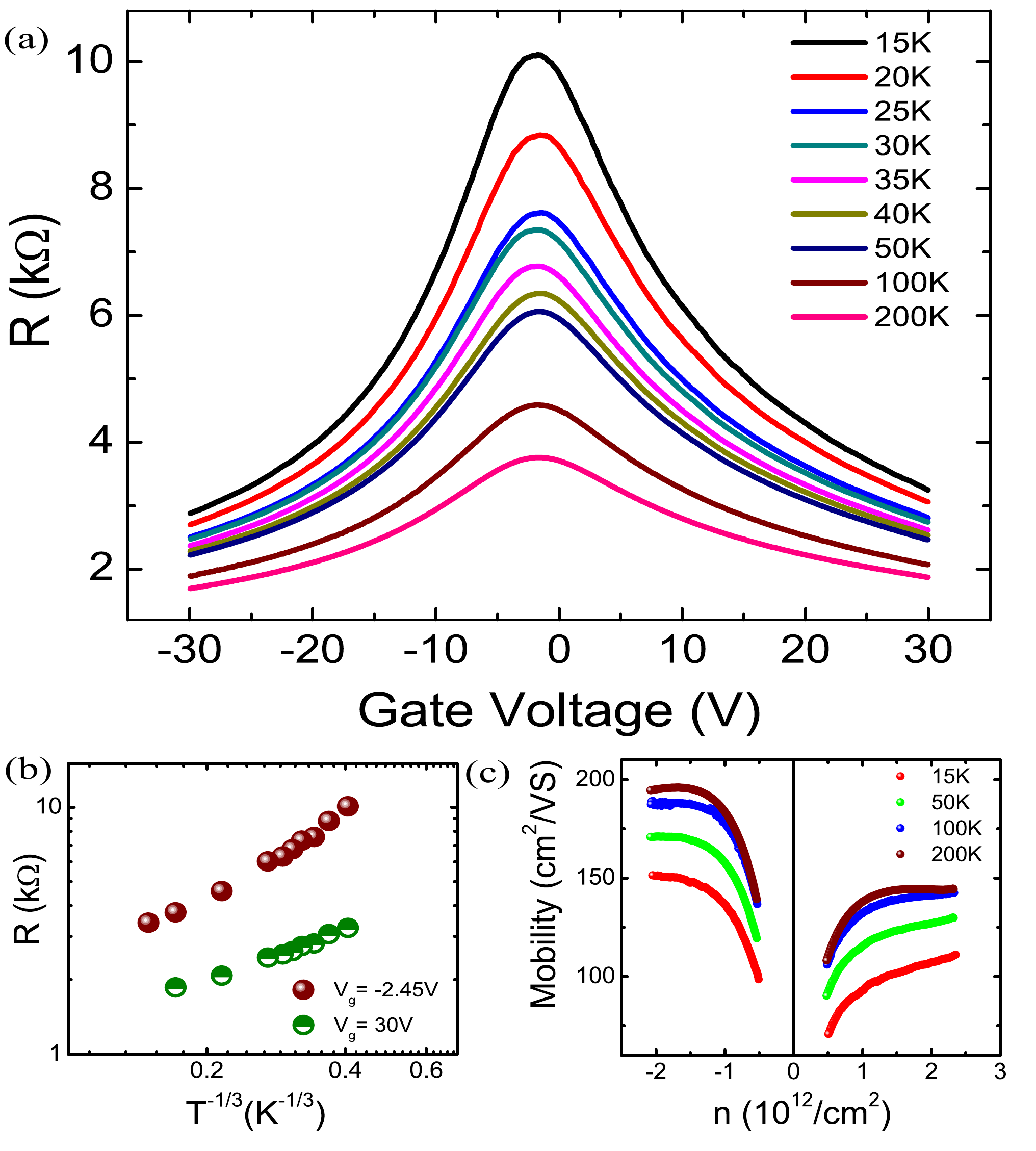}
\end{center}
\caption{Color Online. (a) Resistance vs. gate voltage characteristics for temperature ranging from 15~K to 200~K. (b) Resistance vs. T$^{-1/3}$
are plotted for both V$_g$ = -2.45V (Dirac point) and V$_g$ = 30V (far from the Dirac point), extracted from Fig.~2a. (c) Mobility vs. density
($n$) for various temperature are extracted from Fig.~2a. } \label{figure2}
\end{figure}

The CVD-graphene used in our experiments was grown by decomposition of ethylene on Cu foils at 875~$^0$C as described in Ref~[13]. Then a
polymethyl methacrylate (PMMA) layer was spun on top of the graphene layer formed on the Cu foil, and the Cu foil was then dissolved in 1 M iron
chloride. The remaining graphene/PMMA layer was thoroughly washed with deionized water and transferred to a Si/SiO$_2$ substrate. Subsequently,
the PMMA was dissolved in hot acetone (80~$^0$C) for one hour. The heavily doped silicon was used as backgate. Following transfer to the
Si/SiO$_2$ substrate a detailed Raman spectroscopy was carried out on all our systems. Fig.~1 shows a map of the D-peak ($\sim 1350$~cm$^{-1}$)
intensity from a typical section of our CVD-graphene (Fig.~1a), and two representative spectra (Fig.~1c,d), which indicates a spatially varying
$I_{2D}/I_G$ ratio ($I_{2D}$ and $I_G$ are intensities of the 2D and G bands respectively). Both features can arise from a spatially non-uniform
adhesion/interaction of graphene with the underlying substrate, and associated ripples/local ruptures/line defects/residual byproducts of Cu
etching process etc., highlighting significant disorder of non-Coulombic origin~\cite{APL_QHE}. However, the 2D peaks could be described with a
single Lorentzian line-shape, confirming single-layer graphene. An electrically contacted (with Au metal pads) device is shown in Fig.~1d, where
a five-probe geometry was used to measure the $1/f$-noise in a dynamically balanced Wheatstone bridge configuration. Both standard time-averaged
resistance (utilizing four of the contacts) and noise were measured in low-frequency ac constant current mode. A detail description of the
experimental methods are available elsewhere~\cite{atin_prl,{atin_apl}}. In order to avoid gate leakage-related problems we restricted most
measurements at temperature $T \lesssim 200$~K.

\begin{figure}[!tbp]
\begin{center}
\includegraphics [width=7.5cm,height=8cm]{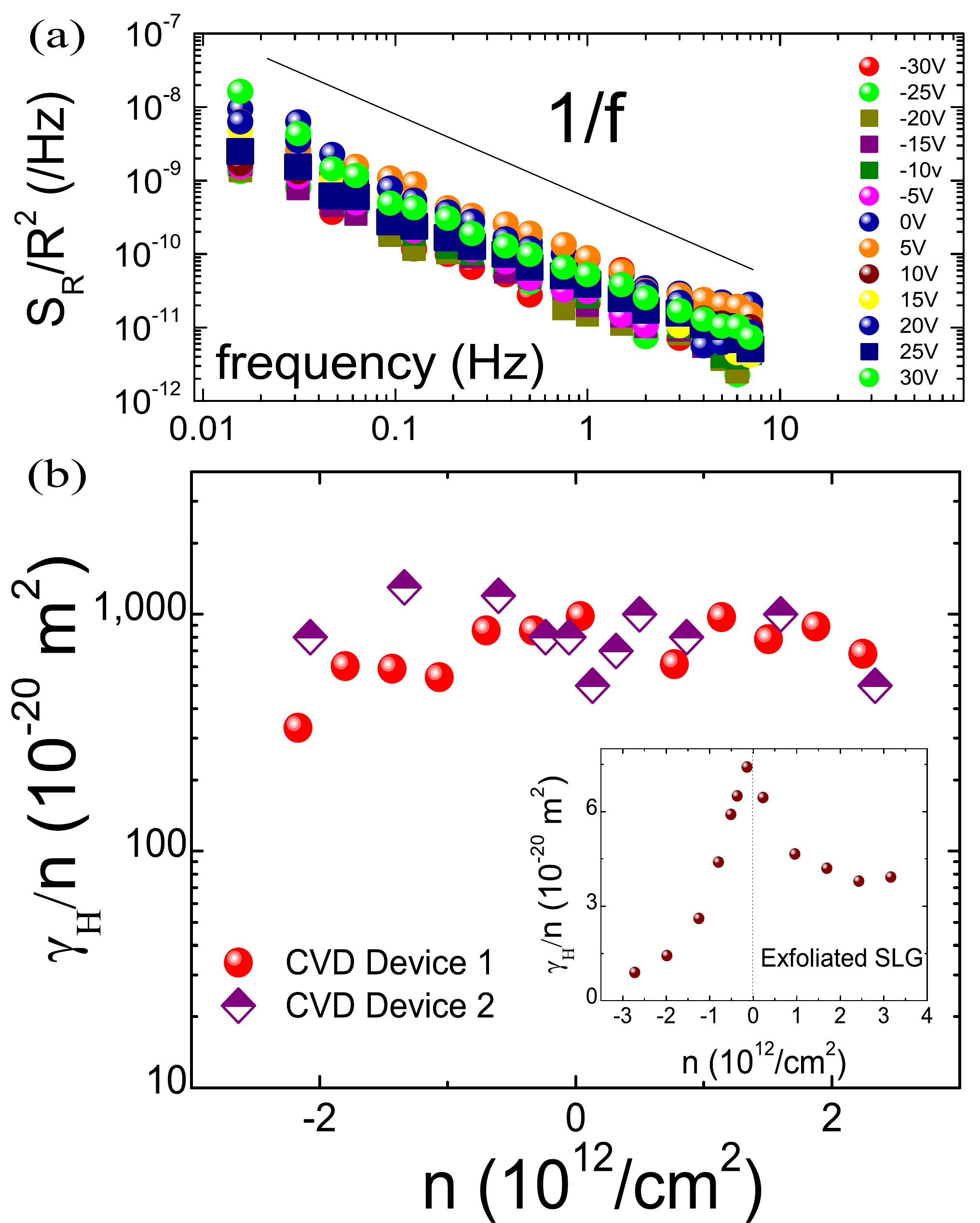}
\end{center}
\caption{Color Online. (a) The noise power spectra $S_R/R^2$ for various gate voltages are shown for T = 100~K, showing $1/f$ type behavior. (b)
Noise magnitude, $\gamma_H/n$ are plotted as a function of density ($n$) for two devices at T = 100~K, showing gate voltage independence of
noise. The inset shows gate voltage dependence of noise magnitude for exfoliated single layer device which shows that noise decreases in both
side of the Dirac point. } \label{figure3}
\end{figure}

The resistance ($R$)$-$backgate voltage ($V_g$) trace of the device in Fig~1d is shown in Fig.~2a for various values of $T$ ranging from 15~K to
200~K. The sheet resistance was found to be $\sim$~$680$~$\Omega$/sq at room temperature. The Dirac point was found to be low, which we believe
to be a combined effect of substrate doping and surface adsorption. $R$ was found to increase sharply with decreasing $T$ at all $V_g$,
reminiscent of the same in ozonization-damaged exfoliated graphene~\cite{Ozonization}. The $T$-dependence of $R$ also seems to indicate a
Mott-type variable range hopping with $\ln R \propto T^{-1/3}$ (see Fig.~2b), although limited range in $T$ or $R$ can make such an analysis
relatively inaccurate.

The noise measurements were performed as function of $V_g$ and $T$. In Fig.~3a
we show the power spectral density (PSD), $S_R$, of noise over nearly three
decades of frequency at various $V_g$ for $T = 100$~K. The PSD can be
normalized with Hooge's phenomenological equation:

\begin{equation}
\label{eq1} S_R = \frac{\gamma_H\langle R\rangle^2}{nA_Gf^\alpha} \
\end{equation}

\noindent where $\gamma_H$ is the Hooge parameter, $A_G$ is the area of
graphene between the voltage probes, and $\alpha$ is the spectral exponent. In
all cases we find $\alpha \approx 1-1.1$, indicating a $1/f$-type spectrum, and
hence, a wide distribution of time scales in the kinetics of disorder. The
noise amplitude, defined as $\gamma_H/n$, was found to be essentially
independent of $V_g$ (or $n$) in both electron and hole-doped regimes.
Different devices showed identical behavior as illustrated in Fig.~3b. This
weak variation in noise, found for all $T$ down to 15~K, is in contrast to the
$n$-dependence of the noise amplitude in exfoliated single layer graphene,
where $\gamma_H/n$ decreases rapidly with increasing $n$ on both sides of the
Dirac point (see inset of Fig.~3b)~\cite{atin_apl}.

\begin{figure}[!tbp]
\begin{center}
\includegraphics [width=1\linewidth]{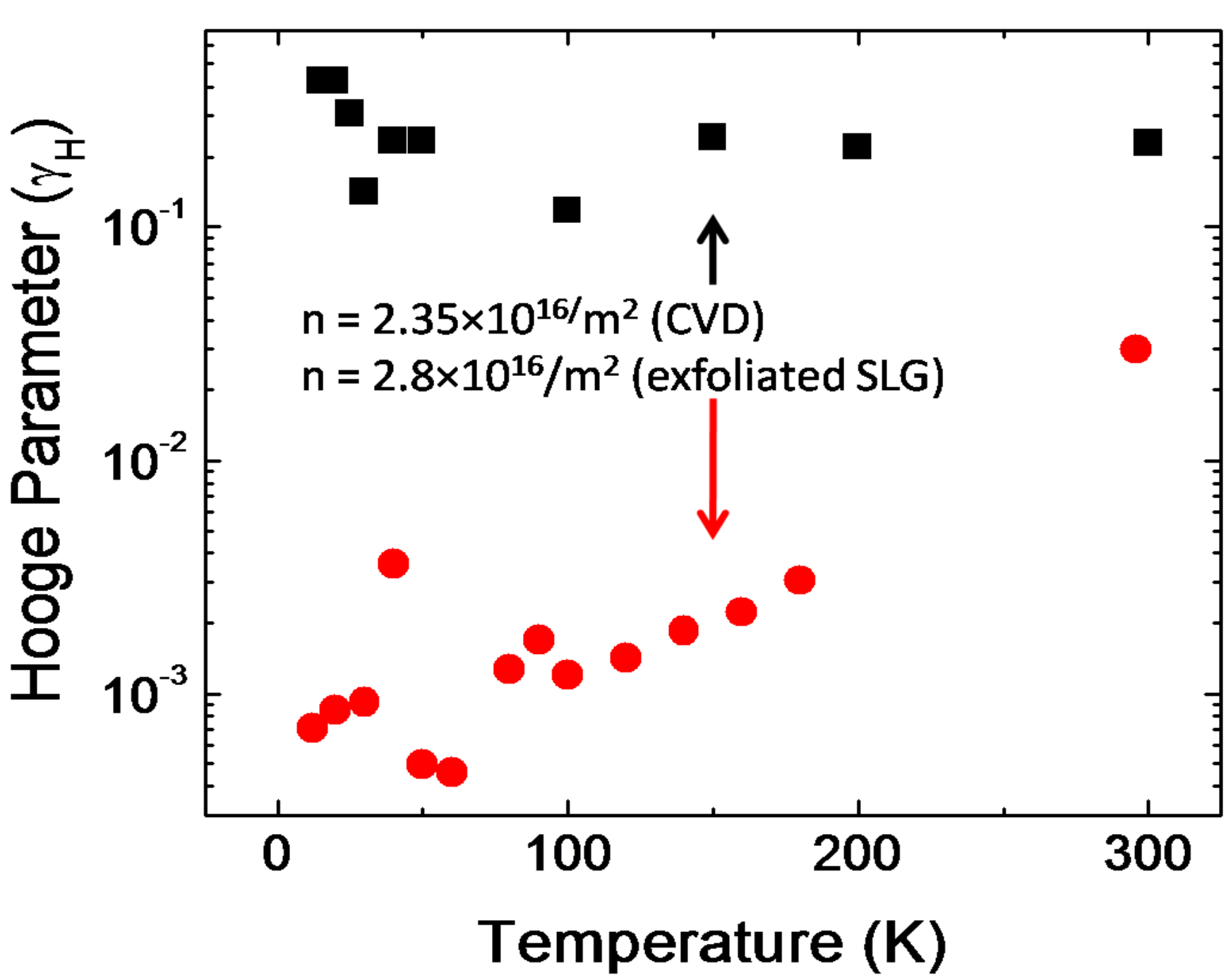}
\end{center}
\caption{Color Online. Comparison of Hooge parameter ($\gamma_H$) at similar carrier densities for both  CVD made graphene and exfoliated
graphene devices over a wide range of temperature (15~K to 300~K). CVD graphene device shows two order of  higher noise magnitude than
exfoliated devices. } \label{figure4}
\end{figure}

Another crucial aspect to note is that the absolute magnitude of $\gamma_H/n$
is nearly hundred times larger in CVD-graphene in comparison to the exfoliated
devices in the same range of $n$. This is highlighted in Fig.~4, where we show
the $T$-variation of $\gamma_H$ (at $n \approx 2.5\times10^{12}$~cm$^{-2}$) for
both graphene systems. For CVD-graphene $\gamma_H$ is not only independent of
$T$ between $\sim 15 - 300$~K, but also displays a $\gamma_H$ that is about one
to two orders of magnitude higher than most graphene based systems~\cite{atin_prl,{atin_apl}}. The difference appears even larger at lower T,
where noise level in exfoliated (or epitaxial) graphene are significantly
reduced.

A possible explanation to the weak variation of noise in CVD-graphene can be through the mechanism of correlated number and mobility
fluctuations due to the trap states at the graphene-substrate interface~\cite{Atin_unpublished,{correlated model_jayaraman}}. Such a mechanism
predicts $S_R \propto 1/n^2$ due to number fluctuations, and $S_R \propto \mu^2$, when mobility fluctuation dominates. In Fig.~2c, we show that
the $\mu$ indeed varies weakly with $n$ in our devices, possibly indicating mobility fluctuations to be the dominant source of noise. However,
similar substrates have been used for exfoliated graphene that showed much lower noise magnitude~\cite{atin_apl}. In our CVD graphene, migration
of surface adsorbates, such as those incurred during the transfer process, or relaxation of structural defects due to the in-built stress may
lead to large noise magnitude.These processes lead to mobility fluctuations, which in an inhomogeneous charge distribution may lead to a gate
voltage (as well as temperature)-independent noise. In the inhomogeneous regime, which can persist upto large $\mid n \mid$ in highly disordered
CVD graphene, the gate voltage is likely to affect relative number of electron and hole puddles rather than the charge density within a
particular puddle significantly~\cite{martin}.

In conclusion, we report experimental investigation of resistance noise in
single layer chemical vapor deposited graphene transferred onto a Si/SiO$_2$
substrate. We find the noise magnitude to be nearly two orders of magnitude
larger than exfoliated single graphene, and largely independent of temperature
and carrier density. A substrate or surface trap-mediated fluctuation model
seems likely, although several details of the noise behavior remains to be
understood quantitatively.

\textbf{Acknowledgement} We acknowledge the Department of Science and Technology (DST) for a funded project, and Indo-US Science and Technology
Forum (IUSSTF) for support. ANP thanks CSIR for financial support.

\end{document}